%% file: main.tex
\documentclass[runningheads]{llncs}
\usepackage[english]{babel}
\usepackage{orcidlink}
\usepackage{ucs}
\usepackage[utf8]{inputenc}
\usepackage[T1]{fontenc}
\usepackage{amssymb}
\usepackage{amsmath}
\usepackage{lmodern}
\usepackage{stmaryrd}
\usepackage{physics} % adds \dd command for differential, like \dd t
\usepackage{array}
\usepackage{textcomp}
\usepackage{uppaal}
\usepackage{graphicx} % Required for inserting images
\usepackage{subcaption}
\usepackage{caption}
\usepackage{booktabs}
\usepackage{todonotes}
\usepackage{siunitx}
\usepackage{booktabs,tabularx}
\newcolumntype{Y}{>{\raggedright\arraybackslash}X}
\usepackage{enumitem}
\usepackage{amssymb}
\usepackage{xcolor}
\usepackage{tikz}
\usepackage{wrapfig}
\usepackage{needspace}
\usetikzlibrary{automata,positioning,arrows.meta}
\usepackage{hyperref}
\hypersetup{
  colorlinks=true,
  urlcolor=blue,
  linkcolor=blue,
  citecolor=blue,
  urlbordercolor=blue,
  pdfborder={1 1 1},
  pdfborderstyle={/S/U/W 1}}

\newcommand{\ok}[1][]{\textcolor{green!50!black}{\checkmark}\if\relax\detokenize{#1}\relax\else\,{\scriptsize(#1)}\fi}
\newcommand{\no}[1][]{\textcolor{red!70!black}{$\times$}\if\relax\detokenize{#1}\relax\else\,{\scriptsize(#1)}\fi}
\newcommand{\field}[1]{\noindent\textbf{#1.}}
\newcommand{\tool}[1]{\texttt{#1}}
\newcommand{\code}[1]{{\uppBasicStyle #1}}

\usepackage{graphicx} % Required for inserting images

\title{UPPAAi: Towards LLM-Assisted Modeling, Verification, Analysis and Synthesis}
\titlerunning{UPPAAi: Towards AI-Assisted Modeling, Verification, and Synthesis}

\author{Peter~Dolog\inst{1}\orcidlink{0000-0003-1842-9131} \and Kim~G.~Larsen\inst{1}\orcidlink{0000-0002-5953-3384} \and Marius~Mikučionis\inst{1}\orcidlink{0000-0001-8157-5428} \and
Frederik~F.~K.~Rasmussen\inst{1}\orcidlink{0009-0000-3817-9489}}
\authorrunning{P. Dolog \and K. G. Larsen \and M. Mikučionis \and F. F. K. Rasmussen}

\institute{Aalborg University, Aalborg, Denmark \email{\{dolog,kgl,marius,ffkr\}@cs.aau.dk}}
\begin{document}

\maketitle
\begin{abstract}
%  \input{abstract}
% \input{abstract_alternativev2}
 \input{abstract3}
\end{abstract}

\section{Introduction}
%\input{intro.tex}
\input{introduction_alternativev2}

\section{Main Features of \uppaal}
\label{sec:uppaal}
\input{upp.tex}

\section{Exposing Tools API}
\label{sec:tools-api}
\input{toolsapi}

\section{Case Studies}
\label{sec:case-studies}
\input{casestudies.tex}

\section{Conclusions}
\label{sec:conclusions}
\input{conclusions}

\clearpage
\bibliographystyle{splncs04}
\bibliography{bib}

\clearpage
\appendix

%\section{Fischer's Mutual Exclusion Protocol}
%\begin{figure}[!tbh]
%  \centering
%  \includegraphics[width=0.4\textwidth]{img/fischer}
%  \caption{Fischer's process model with parameter %\uppConst{pid}.}\label{fig:fischer}
%\end{figure}

\input{demo}

\clearpage
\section{\uppaal Modeling and Query Languages}
\input{ingredients}

\input{queries}

\end{document}

%% file: abstract3.tex
% ---------------------------------------------------------------------------
% Abstract, v2. Included from main.tex inside \begin{abstract}...\end{abstract}.
% ---------------------------------------------------------------------------
For more than three decades, \uppaal{} has supported the modelling and
analysis of real-time and cyber-physical systems.
Yet building a suitable model, formalising the intended properties, choosing
analysis techniques and interpreting counterexamples remain iterative tasks
that demand substantial expertise.
Large language models (LLMs) can offer plausible explanations and even solutions, but they are not necessarily correct.
We present \uppaai{}, an extension of \uppaal{} in which users construct, explain, debug and specify models in dialogue with an LLM.
The guiding principle is that the user requests, the LLM proposes, explains and orchestrates, and \uppaal{} verifies.
\uppaai{} exposes the \uppaal{} editor, queries and verifier through a Tools API.
We report on seven case studies, from explaining to a newcomer to learning a controller, covering timed automata, stochastic hybrid automata and strategy learning.
LLMs proved to be extremely helpful in not just generating solutions but also pro-actively suggesting relevant \uppaal features it was not asked about.
%The main friction lay in the tool interface, such as long-running
%queries that blocked the engine, which points to concrete improvements.
\keywords{Model checking \and Timed automata \and Large language models
\and \uppaal{} \and AI-assisted formal methods}

%% file: introduction_alternativev2.tex
% ---------------------------------------------------------------------------
% Introduction, v2. Included from main.tex after \section{Introduction}.
% Uses \uppaal, \uppaai, \smc, \stratego (uppaal.sty), tikz (main.tex) and
% the section labels sec:uppaal, sec:tools-api, sec:case-study,
% sec:cases-discussion and sec:conclusions.
% ---------------------------------------------------------------------------
\label{sec:intro}

Large language models (LLMs) have become capable generators of software,
producing code, tests and documentation of often remarkable quality from
informal descriptions. Their fundamental weakness is equally well known:
\emph{plausibility is not correctness}. LLMs excel at proposing artefacts that
look right, but are far less reliable at establishing that those artefacts
satisfy subtle concurrency, synchronisation and real-time requirements, which
are precisely the properties that embedded and safety-critical systems depend
on. A growing body of work therefore embeds LLMs in loops in which an external
critic evaluates the generated artefacts and its feedback drives
repair~\cite{beyondVibe2025}, a practice recently termed \emph{loop
engineering}~\cite{loopeng2026}. A key lesson is that such a loop is only as
strong as its critic~\cite{kambhampati2024llmmodulo}: self-evaluation by LLMs
is unreliable~\cite{madaan2023selfrefine}, whereas sound external evaluators
have enabled striking
results~\cite{romeraparedes2024funsearch,trinh2024alphageometry,first2023baldur,wu2024lemur}.

\begin{figure}[t]
  \centering
  \begin{tikzpicture}[
      node distance=12mm and 17mm,
      box/.style={draw, rounded corners, minimum width=20mm,
                  minimum height=8mm, align=center, font=\small},
      lbl/.style={font=\scriptsize, align=center},
      >={Stealth[length=2mm]}]
    \node[box] (user) {User\\(owns requirements)};
    \node[box, right=of user] (llm) {LLM\\(proposes, explains)};
    \node[box, right=of llm] (uppaal) {\uppaal{}\\(verifies)};
    \draw[->] ([yshift=2mm]user.east) -- node[lbl, above]
      {requirements,\\questions} ([yshift=2mm]llm.west);
    \draw[->] ([yshift=-2mm]llm.west) -- node[lbl, below]
      {explanations,\\repairs} ([yshift=-2mm]user.east);
    \draw[->] ([yshift=2mm]llm.east) -- node[lbl, above]
      {models,\\queries} ([yshift=2mm]uppaal.west);
    \draw[->] ([yshift=-2mm]uppaal.west) -- node[lbl, below]
      {verdicts,\\traces} ([yshift=-2mm]llm.east);
    \draw[->, dashed] (user.north) to[bend left=25] node[lbl, above]
      {inspects and edits the model directly} (uppaal.north);
  \end{tikzpicture}
  \caption{The \uppaai{} loop: the LLM generates, the model checker acts as
    the critic, and the user remains in control of the requirements and the
    model.}
  \label{fig:loop}
\end{figure}
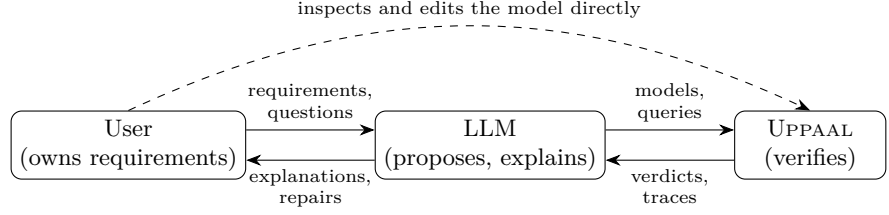

In this paper we instantiate this loop for timed systems with a sound and
exhaustive critic: the model checker \uppaal{}~\cite{behrmann2004tutorial}. We
present \uppaai{}, an extension of \uppaal{} in which a user constructs,
explains, debugs and specifies models in dialogue with an LLM
(Fig.~\ref{fig:loop}). The LLM proposes templates, clocks, invariants,
synchronisations and queries; \uppaal{} checks them; and the resulting
verdicts and counterexamples are explained to the user and drive the next
refinement. Beyond yes/no verification, the same loop covers quantitative
questions through statistical model checking (\uppaal\smc) and the learning of
controllers through \uppaal\stratego~\cite{david2015stratego,larsen2025setss}.
The design follows one principle: \emph{the LLM proposes, explains and orchestrates, but \uppaal{} verifies}.
The LLM contributes language understanding, domain knowledge, abstraction and explanation;
\uppaal{} contributes formal semantics, exhaustive exploration, statistical estimates and counterexamples.
Claims about correctness, performance or synthesised strategies are accepted only when an executable formal or statistical analysis
supports them.

%%%%%%%%%%%%%%%%%%%% Commenting out of the paper
\iffalse
\paragraph{Why a model checker as the critic.}
Our seven case studies (Sect.~\ref{sec:case-studies}) show why the critic
should not be another LLM. In three of them \uppaal{} refuted the assistant:
removing an invariant from Fischer's protocol did not break mutual exclusion
as the assistant predicted, a predicted deadline-miss probability of $1/64$
for a TSN gate schedule turned out to be $2/64$, and two of the assistant's
diagnoses of a verifier that had silently kept an old model were wrong. In a
fourth, an ATM network built from scratch for an engineer new to timed
automata, a correct fix for lost money pushed the worst-case session time from
\SI{197}{s} to \SI{205}{s} and so broke a timing requirement, and scaling from
two to three ATMs exposed a timeout that was safe only for two. Each of these
errors was caught because the claim was put to the engine, not because the
explanation looked implausible. The assistant's role was to formalise the
question and to turn the result into an explanation in domain terms.
\fi
%%%%%%%%%%%%%%%%%% End of commenting out of the paper

\paragraph{Why this matters.}
Each step from informal requirements to a verified model has traditionally
required specialist expertise. With the LLM acting as a \emph{formal-methods
mediator}, an engineer can state a requirement in natural language, obtain a
candidate model and query, and still ask why a clock was introduced or whether
the query captures what was meant. Counterexamples, which traditionally
demanded expertise to interpret, can be explained causally and lead to
proposed repairs that the user can accept or reject. If building a model and
checking a property becomes this cheap, model checking can support
requirements engineering, architecture exploration and teaching, and not only
the late-stage validation of critical components. Finally, as code becomes
cheap, \emph{trusted specifications} become the scarce resource. A validated
\uppaal{} model can serve as a semantic anchor between intent and
implementation: LLM-generated code can be tested for conformance against
it~\cite{larsen2004tron}, or replaced by controllers synthesised from
it~\cite{david2015stratego}. Autonomous software-engineering agents can adopt
the same pattern of generation checked against executable formal semantics as
part of their own reasoning~\cite{dalrymple2024gsai}.

\paragraph{What does not become automatic.}
Verifying a model is not verifying the real system, and an LLM can confidently
construct the wrong abstraction, state a vacuous property, or weaken a query
until it passes. The difficulty thus shifts partly from \emph{checking} to
\emph{model validity}. \uppaai{} addresses this in several ways.
The user can observe, inspect, edit and undo changes in the familiar \uppaal{} editor while the assistant can detect edits the user made in the meantime.
A property can for example only be trusted after it had failed on a deliberately broken variant of the model.
The intended benefit is therefore not only productivity.
Beyond asking whether a property holds, the engineer can ask what the model should be, which properties matter, what a counterexample means, and what follows if a requirement changes.

%%%%%%%%%%%%%%%%% Commenting out of the paper, we described it in general above and we describe in detail in (Sect.~\ref{sec:tools-api})
\iffalse
\paragraph{Integration.}
\uppaai{} exposes the \uppaal{} editor, queries and verifier as a Tools API,
which the LLM can reach in two ways: from an external agentic client (e.g.\
Claude Code) over the Model Context Protocol (MCP), or from a chat embedded in
\uppaal{} itself, which supports several LLM vendors
(Table~\ref{tab:llms}).
The two workflows differ in who owns the conversation and therefore in how context reaches the LLM (Sect.~\ref{sec:tools-api}).
\fi
%%%%%%%%%%%%%%%%% End of Commenting out of the paper

\paragraph{Related approaches.}
LLMs have been coupled with proof assistants~\cite{first2023baldur} and
program verifiers~\cite{wu2024lemur} to generate proofs and verified programs.
There the specification is given and the critic checks the generated
artefact against it. In our setting the LLM also writes the model and the
properties, so the critic checks a model whose fidelity is itself in
question. This is why \uppaai{} keeps the user in charge of the requirements,
makes every change visible in an interactive modelling tool, and asks for the
provenance of each claim to be stated.

\paragraph{Contributions.}
\begin{itemize}
  \item A Tools API that exposes \uppaal{}'s model editing, query
    management and verification to LLMs (Sect.~\ref{sec:tools-api}).
  \item Two integration workflows, an external agentic client via MCP and an
    embedded chat via \texttt{langchain4j}, together with an analysis of
    their trade-offs (Sect.~\ref{sec:tools-api}).
  \item Seven case studies with Claude connected via MCP, starting from
    existing models and from scratch, with users ranging from newcomers to
    experienced modellers, and covering timed automata, stochastic hybrid
    automata, continuous-time Markov chains, ODEs and strategy learning
    (Sect.~\ref{sec:case-studies}).
  \item A cross-case discussion of the recurring patterns: where the verifier
    corrected the assistant, which claims rested on the assistant's argument
    rather than on a tool result, and which limitations of the tool interface
    shaped the work (Sect.~\ref{sec:cases-evaluation}).
\end{itemize}

\paragraph{Outline.}
Section~\ref{sec:uppaal} summarises the features of \uppaal{} for modelling and specification that the LLM must master.
Section~\ref{sec:tools-api} presents the Tools API and the two integration workflows.
Section~\ref{sec:case-studies} reports on the case studies, Section~\ref{sec:cases-evaluation} evaluates and reflects upon the case studies and Section~\ref{sec:conclusions} concludes.

%% file: upp.tex
\paragraph{Modelling.}
\uppaal{} 5.0~\cite{larsen2025setss} unifies, within one modelling language and
one tool, the classical symbolic model checker, \uppaal\smc (statistical
model checking), \uppaal\tiga (timed games) and \uppaal\stratego (timed
Markov decision processes). A model is a network of timed automata,
instantiated from parameterised templates and enriched with discrete data and
C-like code (Table~\ref{tab:ingredients}). The \emph{same} syntactic model can
be given four interpretations, which differ in how non-determinism is resolved.
Each interpretation comes with its own queries and produces its own kind of
result (Table~\ref{tab:formalisms}).

\paragraph{Specifications.}
Symbolic verification uses a fragment of TCTL without nested temporal operators.
It covers \emph{reachability} (\uppProp{E<>}), \emph{safety} (\uppProp{A[]},
\uppProp{E[]}), \emph{liveness} (\uppProp{A<>}, \uppProp{-->}) and deadlock
freedom. Predicates may quantify over process and data ranges with
\uppStat{forall}/\uppStat{exists}. The SMC queries add probabilistic and
expected-value questions over time-, step- or cost-bounded runs. Strategy
queries add synthesis, learning and analysis under a strategy
(\uppStat{under S}). Together these give one coherent language, from yes/no
correctness to quantitative performance and controller synthesis.

%% file: toolsapi.tex
\begin{figure}[tb]
  \centering
  \includegraphics[width=0.85\textwidth]{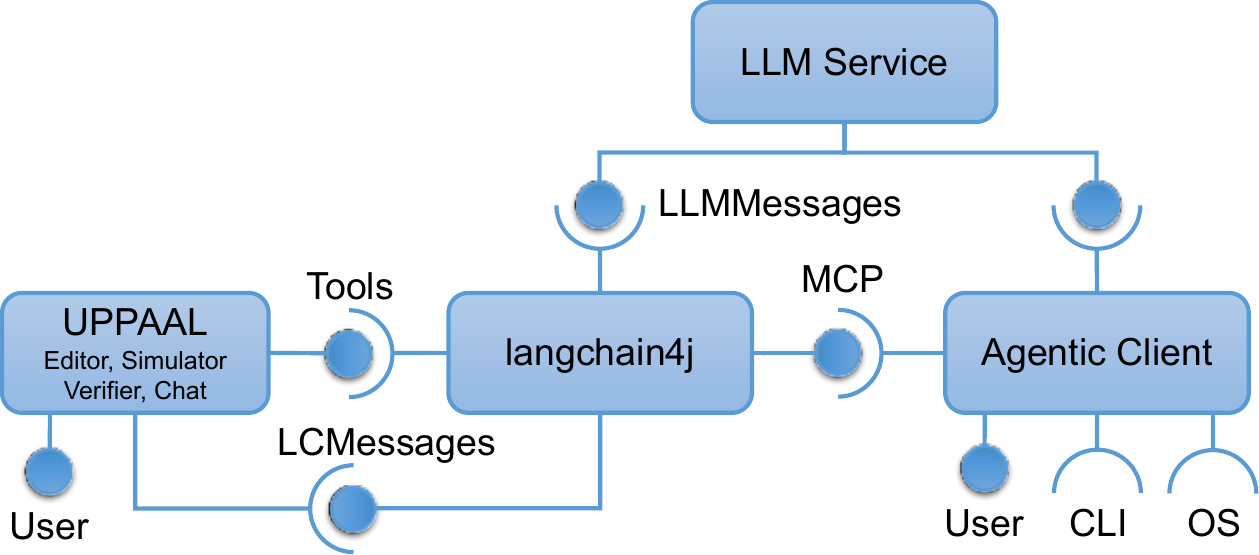}
  \caption{UML component diagram showing \uppaal connecting to LLM Service through the langchain4j library via the LCMessages API, while a vendor-supplied Agentic Client connects via the LLMMessages API and reaches the \uppaal tools through MCP.}\label{fig:interfaces}
\end{figure}

This section describes how \uppaal exposes its functionality to an LLM.
Figure~\ref{fig:interfaces} shows how the software components are connected to provide LLM assistance with \uppaal tools.
Two messaging APIs serve the same purpose of exchanging requests and replies with an LLM, but differ in implementation:
LLMMessages is the vendor's RESTful API accessed over an HTTPS connection, while LCMessages is the Java API of the langchain4j library.

\paragraph{Components}
\begin{itemize}
\item \textbf{LLM Service} exposes the LLMMessages API over HTTPS.
  Access usually requires a separate payment plan and authentication (outside the scope of the diagram).
\item \textbf{Agentic Client} is a vendor-supplied program (e.g., Claude Code) that combines a chat interface with an agent loop and tool calling.
  Such clients are usually marketed as coding assistants, but can be used for any task that their tools support, including modeling.
  It talks to the LLM Service via LLMMessages, acts as an MCP client, and has access to all installed programs via the Command Line Interface (CLI) and the Operating System (OS).
\item \textbf{langchain4j} reaches the LLM Service via LLMMessages (or an equivalent) and offers two interfaces to its consumers: the LCMessages Java API and the Model Context Protocol (MCP), which \uppaal makes available to external clients through a small local HTTP server.
  It does the heavy lifting of messaging and of encoding and decoding the function calls, and uses Java introspection to read the Tools API annotations when compiling the tool description.
\item \textbf{\uppaal} exposes its Tools API as a set of Java functions and consumes the request-response functions of LCMessages.
  It has its own chat history and prompt GUI widgets.
\end{itemize}

\paragraph{Workflow A: external Agentic Client}
The user enters a prompt in the Agentic Client, which owns the chat history.
The RESTful API is stateless, so the client resends the entire history with every request; most clients implement ``history compacting'' to fit the history into the context window and to reduce the input processing costs.
When the LLM Service replies with a tool call, the client forwards it over MCP to langchain4j, which invokes the corresponding \uppaal function and returns the result.
In this workflow \uppaal is passive: it only answers tool calls, and MCP offers no standard way to push changes made to the document by the user (or another LLM) into the model's context.
The client may therefore need to call additional tools (e.g.\ \texttt{getContext}) to find out what the user is looking at and whether the document has been modified.
The context is controlled by the client, not by \uppaal: clients offer their own mechanisms for injecting it, such as instruction files, skills, hooks that run on prompt submission, or user-referenced MCP resources.
In exchange, the user can keep working in a familiar client, which may use additional tools beyond the Tools API, such as the file system, the CLI and the OS.
The Tools API, in contrast, is strictly sandboxed to \uppaal functions: it gives the LLM no access to anything outside the \uppaal model, queries and verifier.

\paragraph{Workflow B: embedded \uppaal chat}
The user enters a prompt in \uppaal, which owns the chat history.
\uppaal uses the LCMessages API of langchain4j to compile a request consisting of the initial prompt instructions, the description of the Tools API, the chat history, additional context information (like what the user has selected or is looking at right now) and lastly the user prompt, which is then passed to the LLM Service.
langchain4j handles the reply from the LLM Service containing instructions to use the tools (call the corresponding functions with certain arguments), marshals the results back to the LLM Service, repeats this as long as further tool calls are requested, and eventually provides the reply to the user, which is appended to the session history.
In this workflow \uppaal is the requesting party and may attach context to every request, so no extra tool calls are needed to learn what the user is looking at.

\paragraph{Comparison}
Both workflows end in the same Tools API.
They differ in who owns the history and initiates requests, and thus in how context reaches the LLM:
Workflow~A reuses an existing, open-ended client but must pull context through tool calls or client-specific injection mechanisms,
while Workflow~B gives \uppaal control over the prompt and context at the cost of a custom chat implementation.

\paragraph{Tools API}
Table~\ref{tab:UPPAAi-tools} lists the tools grouped by purpose: editing the model and its templates, locations, edges and declarations, managing queries, and running and inspecting verification.
Further tools are planned: exporting templates, exposing query descriptions and their customization, a simulation API, and version control integrated with the chat history.
\begin{table}[!bth]
\centering
\caption{Tools API: functions exposed by \uppaal to the LLM.}
\label{tab:UPPAAi-tools}
\footnotesize
\setlength{\tabcolsep}{4pt}
\begin{tabularx}{\textwidth}{@{}l>{\raggedright\arraybackslash}X@{}}
\toprule
\textbf{Function} & \textbf{Description} \\
\midrule
\multicolumn{2}{@{}l}{\emph{Model and templates}} \\
\texttt{getContext}                      & The state of \uppaal GUI (changes and selection) \\
\texttt{get}/\texttt{replaceModel}       & Read or replace the whole model \\
\texttt{checkSyntax}                     & Check the model for syntax errors \\
\texttt{getTemplateList}                 & List the templates in the model \\
\texttt{get}/\texttt{replaceTemplate}    & Read or replace a template \\
\texttt{add}/\texttt{deleteTemplate}     & Add or delete a template \\
\addlinespace
\multicolumn{2}{@{}l}{\emph{Locations and edges}} \\
\texttt{getLocations}                    & List the locations of a template \\
\texttt{add}/\texttt{modify}/\texttt{deleteLocation} & Add, change, or delete a location \\
\texttt{add}/\texttt{modify}/\texttt{deleteEdge}     & Add, change, or delete an edge \\
\texttt{addBranchPoint}                  & Add a branch point for probabilistic edges \\
\texttt{deleteByXPath}                   & Delete any element of a template \\
\addlinespace
\multicolumn{2}{@{}l}{\emph{Declarations}} \\
\texttt{get}/\texttt{setGlobalDeclarations} & Read or replace the global declarations \\
\texttt{get}/\texttt{setLocalDeclarations}  & Read or replace a template's local declarations \\
\texttt{get}/\texttt{setSystemDeclarations} & Read or replace the system declarations \\
\addlinespace
\multicolumn{2}{@{}l}{\emph{Queries}} \\
\texttt{getQueries}                      & List the queries \\
\texttt{insert}/\texttt{removeQuery}     & Add or remove a query \\
\addlinespace
\multicolumn{2}{@{}l}{\emph{Verification}} \\
\texttt{verifyQueries}, \texttt{stop}        & Start or stop verifying queries \\
\texttt{getResults}                          & Read verification progress and results \\
\texttt{getOptions}, \texttt{setQueryOption} & List the engine options, or set one on a query \\
\texttt{getQueryTrace}, \texttt{getQueryPlot}          & Read the trace or plot of a verified query \\
\bottomrule
\end{tabularx}
\end{table}
Table~\ref{tab:llms} shows the LLMs supported by \uppaal.

\begin{table}[tb]
\centering
\caption{Tested support for LLM vendors}
\label{tab:llms}
\footnotesize
\setlength{\tabcolsep}{3pt}
\begin{tabular}{ll}%{\columnwidth}{@{}>{\bfseries}p{2.3cm} >{\raggedright\arraybackslash}X@{}}
\toprule
\textbf{API} & \textbf{Vendors \& Environments} \\
\midrule
MCP & Anthropic Claude Desktop/Code, Mistral Vibe \\
\addlinespace[2pt]
Vendor Specific & OpenAI, Anthropic, Mistral AI \\
OpenAI & OpenAI, DeepSeek, Gemini, Grok, Groq, Ollama \\
\bottomrule
\end{tabular}
\end{table}

%% file: casestudies.tex
\input{cases/cases_intro}

\input{cases/observations}
\subsection{Red/Black Box Sorting on a Conveyor Belt}
\input{cases/box_sorting/summary}
%\input{cases/box_sorting/observations}

%\subsection{Collision-Free Lane Changes}
%\input{cases/car_maneuver/summary}
%\input{cases/car_maneuver/observations}

%\subsection{Spread of COVID-19 in Three Semantics}
%\input{cases/covid19/summary}
%\input{cases/covid19/observations}

\subsection{Storm Water Detention Pond}
\input{cases/storm_water/summary}
%\input{cases/storm_water/observations}

%\subsection{Timing Constraints of an ATM Network}
%\input{cases/atm/summary}
%\input{cases/atm/observations}

\subsection{Robustness of TSN Gate Schedules (IEEE 802.1Qbv)}
\input{cases/tsn/summary}
%\input{cases/tsn/observations}

\Needspace{30\baselineskip}
\subsection{Fischer's Mutual Exclusion Protocol}
\input{cases/fischer/summary}
%\input{cases/fischer/observations}

% ---------------------------------------------------------------------------
% Discussion of the case studies
% ---------------------------------------------------------------------------

\section{Evaluation \& Reflection}
\label{sec:cases-evaluation}
\input{cases/evaluation}

%% file: cases/cases_intro.tex
%Instrument \uppaal to record the time it takes to process the request and make it available in tooltips.
% \ok, \no: optional argument is printed small in parentheses, e.g. \no[6] or \ok[35\,s].

% ---------------------------------------------------------------------------
% Introduction to the case studies
% ---------------------------------------------------------------------------

To assess the capabilities of \uppaai{} we used Claude connected to the \uppaal{} via Model-Context Protocol (MCP) and applied it on seven case studies.
The cases were chosen to cover problems that require
different \uppaal{} modelling formalisms (networks of timed automata,
stochastic timed automata, continuous-time Markov chains, ODEs as hybrid
clocks, and timed games for strategy synthesis and learning with \uppaal\stratego) and
different kinds of queries (safety and reachability, leads-to, deadlock,
\uppProp{sup}/\uppProp{inf} bounds, probability and expectation estimates by
statistical model checking, and strategy learning).

The cases also differ in where the model came from and in what the user
wanted. In three cases an existing \uppaal{} model was the starting point:
Fischer's protocol~\cite{lamport1987fast,behrmann2004tutorial} was to be
explained to a newcomer to model checking; the LEGO MINDSTORMS
sorter~\cite{iversen2000lego} and the lane
changes~\cite{hilscher2011multilane} were to be checked against an informal
requirement and debugged. In the other four cases the assistant created the
model from scratch: the COVID-19 epidemic and the ATM network from Claude's
general knowledge, the storm water detention pond from the user's
requirements, and the TSN gate schedules from a
publication~\cite{craciunas2016qbv} on IEEE 802.1Qbv~\cite{ieee2016qbv}. For
COVID-19 the user deliberately asked for a model built from scratch; when
analysing the results of the session, we consulted an existing \uppaal{}
COVID-19 model~\cite{jensen2020covid}. The users ranged from newcomers to
timed automata to experienced modellers who prescribed modelling decisions.
Table~\ref{tab:cases-overview} summarises the cases.

Each case is reported in the same way: the case and model, the user and
their goal, and the course of the interaction; a table of results per model
variant; the key findings; the user's perspective;
lessons on the tools and the interface;
a table of the \uppaal{} tool calls made in the session;
and metrics of the session. For lack of space we only present summary reports on 4 of the cases spanning  Claude being presented with an existing \uppaal{} model, Claude asked to propose model and query for known  and unknown domains and a case where emphasis is on teaching.

\begin{table}[!bth]
\centering\footnotesize
\caption{Overview of the case studies
%  (source: where the model came from; existing: \protect\uppaal{} model given at the start; general knowledge, requirements,
%  publication: model created from scratch by the assistant from that source; SMC: statistical model checking)
}
\label{tab:cases-overview}
\setlength{\tabcolsep}{3pt}
%\begin{tabularx}{\linewidth}{@{}>{\raggedright\arraybackslash}p{0.19\linewidth}>{\raggedright\arraybackslash}p{0.15\linewidth}>{\raggedright\arraybackslash}X>{\raggedright\arraybackslash}p{0.24\linewidth}@{}}
\begin{tabularx}{\linewidth}{@{}>{\raggedright\arraybackslash}p{0.22\linewidth}>{\raggedright\arraybackslash}p{0.15\linewidth}>{\raggedright\arraybackslash}X>{\raggedright\arraybackslash}p{0.30\linewidth}@{}}
\toprule
\textbf{Case} & \textbf{Source} & \textbf{User's task} & \textbf{Formalism; queries} \\
\midrule
\href{https://deis-tools.github.io/uppaal-models/CaseStudies/AI-IFS2027/fischer/}{Fischer's mutual exclusion protocol}
  & existing \cite{lamport1987fast,behrmann2004tutorial}
  & explore as newcomer; {\scriptsize mutate \& see what breaks}
  & timed automata; safety, deadlock, leads-to\\
\addlinespace
\href{https://deis-tools.github.io/uppaal-models/CaseStudies/AI-IFS2027/box_sorting/}{Box sorting on a conveyor belt}
  & existing~\cite{iversen2000lego}
  & check that boxes are sorted; debug; scale up
  & timed automata, scalar sets; safety, reachability\\
\addlinespace
\href{https://deis-tools.github.io/uppaal-models/CaseStudies/AI-IFS2027/car_maneuver}{Car lane maneuvering}
  & existing~\cite{hilscher2011multilane}
  & fix a model in which two cars collide
  & timed automata with data; reachability, \smc \\
\midrule
\href{https://deis-tools.github.io/uppaal-models/CaseStudies/AI-IFS2027/covid19}{Spread of COVID-19}
  & general knowledge$^{*}$
  & model epidemic from scratch; convert
  & stochastic TA $\rightarrow$ CTMC $\rightarrow$ ODE; \smc \\
\addlinespace
\href{https://deis-tools.github.io/uppaal-models/CaseStudies/AI-IFS2027/storm_water}{Storm water detention pond}
  & requirements
  & synthesise a controller to maintain water level
  & stochastic hybrid game; strategy learning, \smc\\
\addlinespace
\href{https://deis-tools.github.io/uppaal-models/CaseStudies/AI-IFS2027/atm}{Timing constraints in ATM network}
  & general knowledge
  & model as newcomer, verify deadlines
  & timed automata; safety, \uppProp{sup} bounds\\
\addlinespace
\href{https://deis-tools.github.io/uppaal-models/CaseStudies/AI-IFS2027/tsn}{Robustness of TSN gate schedules}
  & publication \cite{craciunas2016qbv,ieee2016qbv}
  & analyse schedules against clock, timing and fault deviations
  & timed automata and an \smc variant; safety, \uppProp{sup}, \smc probabilities\\
\bottomrule
\end{tabularx}
\par\smallskip
\parbox{\linewidth}{\scriptsize $^{*}$When analysing the results of the
session, we consulted an existing \uppaal{} COVID-19
model~\cite{jensen2020covid}.}
\par\smallskip
\parbox{\linewidth}{\scriptsize{Full reports are available at \url{https://deis-tools.github.io/uppaal-models/CaseStudies/AI-IFS2027}}}
\end{table}

%% file: cases/box_sorting/summary.tex
\field{Case \& model} A belt controller in the style of the LEGO MINDSTORMS
sorter~\cite{iversen2000lego}: boxes \uppTemp{RedBox}, \uppTemp{BlackBox} with clock \code{pos},
sensed at \code{pos = 9}, removable in \code{[81,90]}, placed ${\geq}\,10$
apart; on \code{blck}, \uppTemp{TaskMAIN} resets one clock \code{x}, and
\uppTemp{TaskPUSH} ejects at \code{x == 75}.

\field{User \& goal} The modeller asked the assistant (Claude, connected to
\uppaal{} via MCP) whether all black boxes end in \uppLoc{off} and all red
boxes in \uppLoc{end}.

\field{Interaction} Formalise $\to$ refute $\to$ diagnostic queries $\to$
per-box timers $\to$ mutation $\to$ scale up (time-out) $\to$ scalar sets
$\to$ \tool{verifyta} $\to$ re-verify in editor.
\iffalse
\begin{table}[!bt]
\centering\footnotesize
\caption{Results: \protect\ok holds; \protect\no violated;
states: stored by \protect\tool{verifyta}; -- unchecked.}
\label{tab:belt-variants}
\begin{tabular}{@{}lcccccr@{}}
\toprule
Model Variant & timers & boxes B/R & no B at \uppLoc{end} & no R \uppLoc{off} & \code{deadlock} $\rightarrow$ sorted & states\\
\midrule
Original            & 1 & 2/2 & \no  & \ok  & --  & --\\
Per-box timers      & 3 & 3/3 & \ok  & \ok  & \ok & --\\
Mutation: one timer & 1 & 3/3 & \no  & \no  & \no & --\\
Symmetric$^{*}$     & 8 & 5/3 & \ok  & \ok  & \ok & 20\,160\\
Symmetric$^{*}$     & 8 & 8/3 & \ok  & \ok  & \ok & 69\,745\\
Mutation: too few   & 4 & 5/3 & \no  & \no  & \no & 37\,923\\
\bottomrule
\end{tabular}
\par\smallskip
\parbox{\linewidth}{\scriptsize $^{*}$Boxes and timers over \code{scalar}
types; free timer chosen by \code{select}. Intended final state reachable in
all variants checked.}
\end{table}
\fi

\begin{table}[!bt]
\centering\footnotesize
\caption{Results: \protect\ok holds; \protect\no violated;
states: stored by \protect\tool{verifyta}; -- unchecked.}
\label{tab:belt-variants}
\begin{tabular}{@{}lcccccr@{}}
\toprule
Model Variant & timers & boxes & no B at \uppLoc{end} & no R \uppLoc{off} & \code{deadlock} $\rightarrow$ sorted & states\\
\midrule
Original            & 1 & 2B/2R & \no  & \ok  & --  & --\\
Per-box timers      & 3 & 3B/3R & \ok  & \ok  & \ok & --\\
Mutation: one timer & 1 & 3B/3R & \no  & \no  & \no & --\\
Symmetric$^{*}$     & 8 & 5B/3R & \ok  & \ok  & \ok & 20\,160\\
Symmetric$^{*}$     & 8 & 8B/3R & \ok  & \ok  & \ok & 69\,745\\
Mutation: too few   & 4 & 5B/3R & \no  & \no  & \no & 37\,923\\
\bottomrule
\end{tabular}
\par\smallskip
\parbox{\linewidth}{\scriptsize $^{*}$Boxes and timers over \code{scalar}
types; free timer chosen by \code{select}. Intended final state reachable in
all variants checked.}
\end{table}

\field{Key findings}
\begin{enumerate}[label=F\arabic*.,leftmargin=2em,itemsep=0pt,topsep=1pt]
\item \emph{A single timer loses ejects.} Instead of a trace, Claude asked
  separating reachability queries: \code{E<> B1.end \&\& B2.start} is
  unreachable, \code{E<> B1.end \&\& B2.off} reachable. A second black box
  resets \code{x}, and the first passes the piston (transit 75 $\gg$ spacing
  10). The explanation is Claude's; the verdicts are \uppaal{}'s.
\item \emph{The checks were checked.} With one timer, and with four timers
  for five boxes, \uppaal{} reports the properties violated, so they are not
  vacuous. Unexplained side effect: a red box can then be removed too.
\item \emph{Scaling needed remodelling.} Five black boxes as named instances
  exceeded the time-out; scalar sets enabled symmetry reduction, after
  replacing an ordered \code{firstFree()} that broke symmetry (5 boxes:
  3\,s).
\end{enumerate}

\field{User's perspective} The informal requirement became the safety
property
\begin{uppaalcode}
A[] deadlock imply ((forall (i : bid_t) BlackBox(i).off) &&
                    (forall (j : rid_t) RedBox(j).end))
\end{uppaalcode}
\noindent i.e.\ whenever the system is at rest, every box is sorted;
``eventually sorted'' fails only because the guard may idle. The user chose
the scale and stopped the stuck verification.

\field{Tool \& interface lessons} (i)~A timed-out verification kept running
and blocked every request, even after a GUI stop. (ii)~Fallback to
\tool{verifyta} of the same version; provenance had to be stated.
(iii)~Traces not requested; diagnosis by queries. (iv)~``8 timers suffice''
is an argument, not a proof.

\begin{table}[!tb]
\centering\footnotesize
\caption{\protect\uppaal{} MCP tool calls in the session.}
\label{tab:belt-tools}
\begin{tabularx}{\linewidth}{@{}l>{\raggedright\arraybackslash}Xr@{}}
\toprule
Purpose & Tools (calls) & Total\\
\midrule
Inspect & \tool{getContext} (5), \tool{getModel} (2), \tool{getQueries} (1),
  \tool{getTemplate} (1) & 9\\
Verify  & \tool{verifyQueries} (7) & 7\\
Queries & \tool{insertQuery} (18), \tool{removeQuery} (12) & 30\\
Edit    & \tool{setGlobalDeclarations} (4), \tool{checkSyntax} (4),
  \tool{replaceTemplate} (2), \tool{setSystemDefinition} (2),
  \tool{replaceModel} (1) & 13\\
\midrule
All     & 12 distinct tools & 59\\
\bottomrule
\end{tabularx}
\end{table}

\field{Metrics} 4 prompts (+1 mid-turn note); 59 MCP tool calls, 9 failed
(1 time-out, 8 ``busy''); 6 verification runs in the editor (28 query
checks) and 5 \tool{verifyta} runs (25 checks, ${\leq}\,14$\,s); 6 variants;
0 traces requested; 1 unexplained observation; 17\,min wall clock, about
8\,min assistant time.

% Session files: \url{https://deis-tools.github.io/uppaal-models/CaseStudies/AI-IFS2027/box_sorting}

%% file: cases/storm_water/summary.tex
\field{Case \& model} \uppaal\stratego model built from scratch:
\uppTemp{Rain} (CTMC over 5 rain classes, 4 scenarios, exponential
durations), \uppTemp{Pond} (level \code{w} as hybrid clock: runoff in,
orifice outlet, weir spillway above 2.5\,m, evaporation; 4 monitor clocks),
\uppTemp{Controller} (every 30\,min sets outlet
\code{u}\,$\in\{0,\tfrac13,\tfrac23,1\}$).

\field{User \& goal} A user asked the assistant (Claude, connected to
\uppaal{} via MCP) for a model with rain scenarios and a learned periodic
controller keeping the level between min (1.0\,m) and max (2.5\,m): no
overflow, average level as high as possible.

\field{Interaction} Build $\to$ learn \code{minE(dist + 10$^4$excess +
100deficit)} $\to$ recalibrate outlet $\to$ relearn $\to$ save/load strategy
$\to$ explain \uppTemp{Rain}, \uppTemp{Pond} $\to$ speed up learning with
\tool{verifyta} $\to$ write-ups.

\begin{table}[!tb]
\centering\footnotesize
\caption{Results per model variant over 24\,h, random scenario (1000 runs,
95\,\% CI; $Q_{\max}$: fully open outlet flow in m$^3$/min; severity in
m$\cdot$min; --: not checked).}
\label{tab:pond-variants}
\setlength{\tabcolsep}{4pt}
\begin{tabular}{@{}lccccc@{}}
\toprule
Variant & $Q_{\max}$ & avg level & P(overflow) & severity & P(below min)\\
\midrule
random              & 60  & 0.47\,m          & $6.4\pm4.0\,\%$     & --   & --\\
\code{Opt}          & 60  & 1.08\,m          & $3.6\pm3.4\,\%$     & 0.30 & 94\,\%\\
random              & 120 & 0.22\,m          & ${\leq}\,5\,\%$     & --   & --\\
\code{Opt}          & 120 & \textbf{1.32\,m} & $5.5\pm3.9\,\%$     & 0.06 & 58\,\%\\
\code{Opt} reloaded & 120 & 1.31\,m          & --                  & --   & --\\
\bottomrule
\end{tabular}
\end{table}
%\vspace{-6pt}

\field{Key findings}
\begin{enumerate}[label=F\arabic*.,leftmargin=2em,itemsep=0pt,topsep=1pt]
\item \emph{A timid controller pointed to the plant.} With $Q_{\max}=60$ the
  strategy hugged the minimum: extreme rain (180\,m$^3$/min) exceeds the
  open outlet, so overflow was unavoidable. Doubling the outlet, not tuning
  the learner, raised the level by 0.24\,m and cut overflow severity
  5$\times$.
\item \emph{Safe by being empty.} Random control overflows as rarely as
  \code{Opt}, but at 0.22\,m; safety alone says little.
\item \emph{The strategy lived outside the model.} On re-running queries the
  user found \code{Opt} gone: strategies stay in engine memory, never in the
  model file. Two queries, \code{saveStrategy} and \code{loadStrategy},
  fixed it.
\item \emph{Time went into the ODE, not the learner.} Discretisation step
  0.01\,$\to$\,0.5\,min cut learning from ${\approx}\,7$ to ${<}\,2$\,min at
  equal quality; a smaller learning budget instead gave 28$\times$ the
  overflow severity.
\end{enumerate}

\field{User's perspective} Questions concerned artefacts and cost, not
syntax: ``why was \code{Opt} not saved?'', ``10 minutes is a long time to
wait'', ``which query gives the average level?''; explanations worked when
grounded in numbers (rain shares, steady-state levels). The assistant
flagged its own rain data as stress tests (34--108\,mm/day vs.\
${\approx}\,2$ in Denmark).

\field{Tool \& interface lessons} (i)~Synchronous calls time out on
minute-long learning, and a busy engine blocks every call, even
\tool{getModel}. (ii)~No access to engine options; \tool{verifyta} was run
directly. (iii)~\tool{replaceModel} loses undo history: the user's unsaved
model was backed up first. (iv)~\tool{getModel} returns stored simulation
data (${\approx}\,380$\,KB). (v)~\code{trans} and \code{under} are reserved
words.

\begin{table}[!tb]
\centering\footnotesize
\caption{\protect\uppaal{} MCP tool calls in the session.}
\label{tab:pond-tools}
\begin{tabularx}{\linewidth}{@{}l>{\raggedright\arraybackslash}Xr@{}}
\toprule
Purpose & Tools (calls) & Total\\
\midrule
Inspect & \tool{getModel} (5), \tool{getQueries} (4), \tool{getContext} (1) & 10\\
Verify  & \tool{verifyQueries} (7) & 7\\
Queries & \tool{insertQuery} (8), \tool{removeQuery} (6) & 14\\
Edit    & \tool{replaceModel} (2), \tool{setGlobalDeclarations} (2),
  \tool{checkSyntax} (3), \tool{modifyLocation} (1) & 8\\
\midrule
All     & 10 distinct tools & 39\\
\bottomrule
\end{tabularx}
\end{table}
%\vspace{-6pt}

\field{Metrics} 12 prompts; 39 MCP tool calls, 8 without a usable result
(2 time-outs, 4 ``busy'', 2 keyword errors); 3 GUI learnings (2 by the
assistant, ${\approx}\,7$\,min each; 1 by the user) + 4 learning and 4
simulation-timing runs with \tool{verifyta}; 2 plant variants; 12 queries;
1 strategy file (34\,KB).

% Session files: \url{https://deis-tools.github.io/uppaal-models/CaseStudies/AI-IFS2027/storm_water}

%% file: cases/tsn/summary.tex
\field{Case \& model} Gate schedules for time-sensitive networking
(Craciunas et al.~\cite{craciunas2016qbv}), checked for robustness against worse clocks, late
releases, frame loss and best-effort traffic. A small network (two talkers,
a switch, a listener) with three schedules derived from the paper's
constraints; ten templates -- clocks, sources, gates, schedulers, links,
receiver, best-effort source, fault environment, passive observer -- plus an
\smc{} variant.

\field{User \& goal} A modeller asked the assistant (Claude, connected to
\uppaal{} via MCP and \tool{verifyta}) to model and analyse this robustness,
prescribing explicit transmission durations, stated traffic and fault
assumptions, correlated clock uncertainty, packet identity and passive
observers.

\field{Interaction} Model $\to$ verify $\to$ sweep 29 configurations $\to$
counterexamples $\to$ \smc{} (three variant defects fixed) $\to$ template
walk-through and counterexample timelines $\to$ \smc{} queries in the model
$\to$ review of timing uncertainties.

\begin{table}[!bth]
\centering\scriptsize
\caption{Results per model variant (\protect\ok: holds; \protect\no:
violated; WCRT: exact worst-case latency; Pr: \protect\smc{} probability of
a deadline miss within 10 cycles; cex: steps of the shortest counterexample;
--: not checked).}
\label{tab:tsn-variants}
\setlength{\tabcolsep}{2.8pt}
\begin{tabular}{@{}lcccccr@{}}
\toprule
Variant & deadline & window & overrun & WCRT & Pr & cex\\
\midrule
CFG0 (shared window), $\Delta\leq 2$                   & \ok & \ok       & \ok & $46{+}\Delta$ & 0              & --\\
CFG0, $\Delta=3$ (drifting offsets)                    & \no & \ok$^{*}$ & \ok & 133 (139)     & 0.032 (0.14)   & 51\\
CFG0, release jitter 11                                & \no & \ok$^{*}$ & \ok & 148           & 0.85           & 34\\
CFG1 (no isolation), $\Delta=0$                        & \no & \no       & \ok & 46            & 0.50$^{\S}$    & 54\\
CFG2 (time isolation), $\Delta\leq 2$; ${\leq}\,3$ losses & \ok & \ok    & \ok & 38            & --             & --\\
BE, length-aware gates, or guard band 8                & \ok & \ok       & \ok & 48            & 0              & --\\
BE, no length check, guard band 0/4/7                  & \no & \ok       & \no & 56/52/49      & .72/.16/.002   & 51\\
naive per-event jitter, $\Delta=0$                     & \no & --        & --  & --            & --             & 32\\
\bottomrule
\end{tabular}
\par\smallskip
\parbox{\linewidth}{\scriptsize $^{*}$Window indices only; the frame pushed
into the next cycle is caught by the deadline. $^{\S}$Within one cycle.}
\end{table}
%\vspace{-5mm}

\field{Key findings}
\begin{enumerate}[label=F\arabic*.,leftmargin=2em,itemsep=0pt,topsep=1pt]
\item \emph{The model checker corrected the assistant.} Claude predicted a
  miss probability of $1/64$ at $\Delta=3$; \smc{} gave $[0.027,0.038]$, and
  the exhaustive engine, asked per offset combination, found a second
  failing case (a late switch: latency $49>48$), so $2/64$.
\item \emph{The two engines checked each other.} Three \smc{} results
  contradicted verdicts or predictions, each caused by Claude's \smc{}
  variant (a rate with a clock guard never fired; an equality-guarded choice
  was biased).
\item \emph{The clock model decides the verdict.} Correlated per-node
  offsets are robust up to the synthesised $\delta=2$; independent per-event
  jitter reports violations even for perfect clocks.
\item \emph{Isolation and guard bands are necessary.} Without isolation the
  schedule fails with perfect clocks; without length checks only a guard
  band of one maximal BE frame helps, and guard band 7 leaves a rare event
  ($3/2000$ runs) found exhaustively in 0.5\,s.
\end{enumerate}

\field{User's perspective} Each prescribed modelling decision showed up in a
result; the timing review exposed what remains exact (whole-tick offsets,
stepwise drift, fixed transmission and forwarding times).

\field{Tool \& interface lessons} (i)~MCP verifies only the editor's model,
so sweeps ran through \tool{verifyta}. (ii)~An unsaved model in the editor
was backed up before \tool{replaceModel}. (iii)~Traces are off by default.
(iv)~\smc{} needs broadcast channels and its rate and point-guard semantics
must be cross-checked.

\begin{table}[!htb]
\centering\footnotesize
\caption{\protect\uppaal{} MCP tool calls in the session.}
\label{tab:tsn-tools}
\begin{tabularx}{\linewidth}{@{}l>{\raggedright\arraybackslash}Xr@{}}
\toprule
Purpose & Tools (calls) & Total\\
\midrule
Inspect & \tool{getContext} (2), \tool{getModel} (2),
  \tool{getTemplateList} (1), \tool{getQueries} (1) & 6\\
Verify  & \tool{verifyQueries} (2) & 2\\
Edit    & \tool{replaceModel} (2), \tool{checkSyntax} (3) & 5\\
\midrule
All     & 7 distinct tools & 13\\
\bottomrule
\end{tabularx}
\end{table}
%\vspace{-5mm}

\field{Metrics} 8 prompts; 13 MCP tool calls, 0 failed; about 115
\tool{verifyta} runs (statistics sweep 48\,min CPU, largest run
$7\cdot10^6$ states, 1.3\,GB); 6 counterexamples (32--54 steps); about 60
\smc{} estimates; 1 wrong prediction and 3 variant defects caught.

% Session files: \url{https://deis-tools.github.io/uppaal-models/CaseStudies/AI-IFS2027/tsn}

%% file: cases/fischer/summary.tex
\field{Case \& model} Fischer's protocol~\cite{lamport1987fast} (\uppaal{}

\begin{wrapfigure}[6]{R}{0.36\textwidth}
  \vspace{-22mm}
  \includegraphics[width=\linewidth]{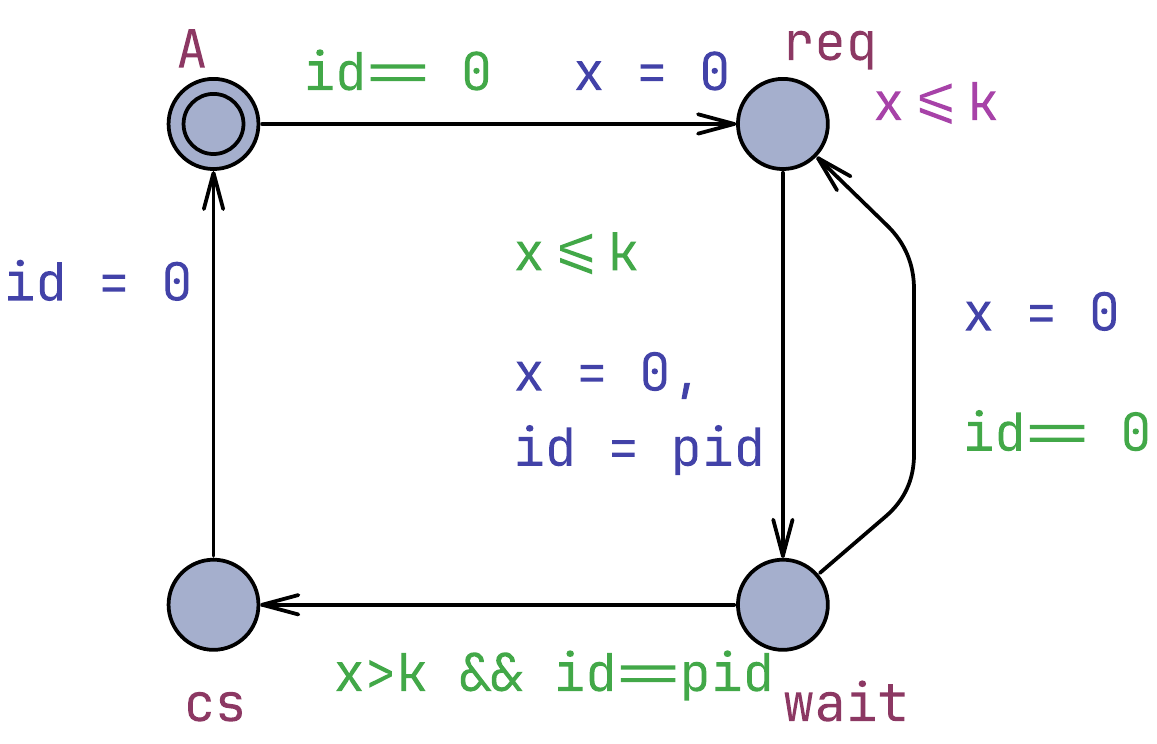}
  \caption{Fischer process template \uppProc{P} with parameter \uppConst{pid}.}\label{fig:fischer}
\end{wrapfigure}

\noindent demo~\cite{behrmann2004tutorial}, Fig.~\ref{fig:fischer}): six instances of
\uppTemp{P}, shared \uppVar{id}, clock \uppClock{x}, \uppConst{k}${}=2$;
\uppVar{id} is written with \uppConst{pid} within \uppInv{x<=k} and checked when \uppGuard{x>k}.%

\field{User \& goal} A user new to model checking asked the assistant
(Claude, connected to \uppaal{} via MCP) to explain the model open in the
editor.

\field{Interaction} Explain $\to$ verify $\to$ two mutations $\to$ fairness
query $\to$ starvation refinement $\to$ restore $\to$ simulator walk-through.

\field{Key findings}
\begin{enumerate}[label=F\arabic*.,leftmargin=2em,itemsep=0pt,topsep=1pt]
\item \emph{The model checker corrected the assistant.} It predicted that
  removing \uppInv{x<=k} on \uppLoc{req} would break mutual exclusion; it
  held (the edge guard remains), while deadlock freedom and
  \uppLoc{req}$\leadsto$\uppLoc{wait} failed
  (Table~\ref{tab:fischer-variants}).
\item \emph{Changing one comparison exposed a race at a single instant.} With
  \uppGuard{x>=k}, a check at exactly $x=k$ meets a still-permitted write
  (6 steps, 0.007\,s), an instant random simulation rarely hits.
\item \emph{The first liveness counterexample was not the intended one.}
  \uppLoc{req}$\leadsto$\uppLoc{cs} failed because the \emph{winning}
  \uppProc{P(1)} may idle in \uppLoc{wait}; starvation needed a refinement.
\end{enumerate}

\vspace*{-7mm}
\begin{table}[!h]
\centering\footnotesize
\caption{Results: \protect\ok holds; \protect\no violated,
with counterexample length; -- unchecked.}
\label{tab:fischer-variants}
\begin{tabular}{@{}lcccc@{}}
\toprule
Model Variant & mutex & no deadlock & \uppLoc{req}$\leadsto$\uppLoc{wait} & \uppLoc{req}$\leadsto$\uppLoc{cs}\\
\midrule
Original                                 & \ok    & \ok    & \ok    & \no[4, ``lazy'']\\
\uppGuard{x > k} $\to$ \uppGuard{x >= k} & \no[6] & --     & --     & --\\
no invariant on \uppLoc{req}             & \ok    & \no[6] & \no[1] & --\\
eager variant$^{*}$                      & \ok    & \ok    & \ok    & \no[15, lasso]\\
\bottomrule
\end{tabular}
\par\smallskip
\parbox{\linewidth}{\scriptsize $^{*}$Invariants on \uppLoc{wait},
\uppLoc{cs}; a \uppLoc{lost} location; an \uppStat{urgent broadcast} channel
forcing retries.}
\end{table}
\vspace*{-5mm}

\field{User's perspective} \emph{Guards allow, invariants force}; a leads-to
trace that \emph{stops} means ``may stay here forever''; breaking taught more
than ``satisfied''.

\field{Tool \& interface lessons} (i)~No GUI or simulator access.
(ii)~\tool{getContext} detected user edits before restoring.
(iii)~\tool{replaceModel} loses undo history. (iv)~Traces off by default.

\begin{table}[!b]
\centering\footnotesize
\caption{\protect\uppaal{} MCP tool calls in the session.}
\label{tab:fischer-tools}
\begin{tabularx}{\linewidth}{@{}l>{\raggedright\arraybackslash}Xr@{}}
\toprule
Purpose & Tools (calls) & Total\\
\midrule
Inspect & \tool{getModel} (3), \tool{getQueries} (2), \tool{getContext} (2),
  \tool{getVerificationOptions} (1) & 8\\
Verify  & \tool{verifyQueries} (11), \tool{getQueryTrace} (5) & 16\\
Queries & \tool{setQueryOption} (9), \tool{insertQuery} (1),
  \tool{removeQuery} (1) & 11\\
Edit    & \tool{modifyEdge} (5), \tool{modifyLocation} (4), \tool{addEdge} (2),
  \tool{addLocation} (1), \tool{setGlobalDeclarations} (1),
  \tool{replaceModel} (1), \tool{checkSyntax} (1) & 15\\
\midrule
All     & 16 distinct tools & 50\\
\bottomrule
\end{tabularx}
\end{table}

\field{Metrics} 11 prompts (4 picks from the assistant's menu); 50 MCP tool
calls, 0 failed; 11 verification runs (27 query checks, ${\leq}\,2.1$\,s);
4 variants; 5 counterexamples (1--15 steps); 1 wrong prediction caught;
4 restores, all re-verified.

%Session files: \url{https://deis-tools.github.io/uppaal-models/CaseStudies/AI-IFS2027/fischer}

%% file: cases/evaluation.tex
% Requires in the preamble: \usepackage{booktabs,tabularx}

The cases fall into four groups (Table~\ref{tab:sessions}): debugging existing
models, modelling in domains known or new to the user, and teaching.

\begin{table}[!h]
\caption{The sessions side by side, emphasising acceleration (Pr.: prompts;
Calls: MCP tool calls; Beyond the request: contributions the user did not ask
for; Reference: what had been achieved, or what the user knew, without the LLM).}
\label{tab:sessions}
\scriptsize
\setlength{\tabcolsep}{3pt}
\begin{tabularx}{\textwidth}{@{}l r r >{\raggedright\arraybackslash}X >{\raggedright\arraybackslash}X >{\raggedright\arraybackslash}X@{}}
\toprule
Case & Pr. & Calls & Obtained with UPPAAi & Beyond the request & Reference \\
\midrule
\multicolumn{6}{@{}l}{\emph{Debugging an existing model}}\\
Belt & 4 & 59
  & bug suspected at once and confirmed by UPPAAL; explained from the trace; fixed with one timer per black box; final model verifies in seconds
  & bound of at most 8 boxes on the belt; state-space explosion diagnosed; scalar sets for symmetry reduction
  & model never taken this far by us; session of 17\,min \\
Lanes & 12 & 56
  & two bugs found with UPPAAL, explained and fixed; SMC guarantees for the full model with more lanes and cars, in a full sweep
  & lanes shortened (700\,$\to$\,120) to fit the state space
  & buggy left-over; unresolved after repeated half-hour attempts \\
\midrule
\multicolumn{6}{@{}l}{\emph{Known domain}}\\
COVID & 6 & 51
  & agent-based model (one automaton per individual), fully parameterised; SMC queries in correct syntax; CTMC and ODE variants
  & clean-up of previous session; extensions (age structure, threshold-triggered lockdown or vaccination, hospital capacity); consistency of the three models shown
  & agent-based model was the end goal of our earlier work~\cite{jensen2020covid} \\
Pond & 12 & 39
  & \textsc{Stratego} model with randomised rain scenarios and ODE pond dynamics; learned periodic outlet controller balancing overflow against average level
  & previous model saved and cleared; step-size study with \textsc{Stratego} to shorten learning
  & complete model and controller in one session \\
\midrule
\multicolumn{6}{@{}l}{\emph{New domain}}\\
TSN & 8 & 13
  & model of gate schedules from a paper using other methods, on a user-suggested architecture; experiments and SMC queries answered by UPPAAL
  & experiments proposed from protocol knowledge; roles of templates explained; refuted predictions explained
  & user knew real-time protocols, not TSN; only a paper given \\
ATM & 10 & 70
  & network of timed automata with deadline queries from scratch; scaled from two to three ATMs
  & --
  & engineer new to timed automata \\
\midrule
\multicolumn{6}{@{}l}{\emph{Teaching}}\\
Fischer & 11 & 50
  & model recognised and explained in domain terms; mutations with counterexamples; 15-step starvation loop; simulator prepared
  & numbered menu of four experiments; model restored after each; own omission noticed and repaired
  & learner new to model checking; little effort needed \\
\bottomrule
\end{tabularx}
\end{table}

\paragraph{Debugging existing models.}
In the belt case Claude immediately suspected a bug, had \uppaal confirm it,
explained it from the trace and fixed it with one timer per black box. It
argued correctly that at most eight boxes fit on the belt, and, when
verification stalled, proposed scalar sets for symmetry reduction; the final
model verifies in seconds. We never took this model this far ourselves. The
lane-change model was a buggy left-over that we had repeatedly failed to fix
in spare half-hours. Claude found two bugs, explained and fixed them, shortened
the lanes (700\,$\to$\,120) to tame the state space, and, at the user's
suggestion, swept the full model with SMC.

\paragraph{Known domains.}
For COVID-19, Claude's first proposal was an agent-based model, the very model
our earlier work~\cite{jensen2020covid} had aimed at. It was flawless and fully
parameterised, and came with SMC queries in impeccable syntax. Claude then
provided CTMC and ODE variants and showed that the three models are
consistent. For the detention pond it built a \stratego model with
randomised rain scenarios and a learned outlet controller. When learning
became slow, Claude used \stratego itself to find a suitable ODE step
size.

\paragraph{New domains.}
For TSN, the user knew real-time protocols but not TSN. Claude received only
a paper~\cite{craciunas2016qbv} using different methods and an architecture
suggested by the user. The resulting model appears to be of high quality;
Claude explained each template's role and proposed experiments and SMC
queries. Where \uppaal refuted its predictions, Claude explained why. For the
ATM network, an engineer new to timed automata obtained a verified model from
scratch.

\paragraph{Teaching.}
Claude recognised Fischer's protocol, explained it in domain terms and offered
four numbered experiments. It mutated and restored the model, and accepted
\uppaal's refutation of its own prediction. It produced a starvation loop and
prepared the simulator for a counterexample, which was very educational with
little effort from the learner.

\paragraph{Potential pitfalls/misunderstandings.}
Newcomers often ask brief and generic questions to which LLMs answer in generic terms with superficial tailoring towards their specific use-case. As a result such proposals contain unchecked speculative information (user did not ask to check so no checks are performed at a time) and then later it may come out as a surprise that generic predictions did not hold up in a particular case.

\paragraph{Potential information leaks.}
Local MCP server is accessible only by running AI agents locally and thus giving them access to all local resources. 
This instrumentation makes AI agents very powerful, but may pose a risk of information leak.
For example, in TSN case Claude Code proactively searched the computer for \texttt{verifyta} and then used it to sweep model parameters outside \uppaal workspace.
Alternatively, \uppaai supports a number of remote LLM APIs and thus limits the AI agent access to its Tools API, making it a safer alternative. However, many vendors require a separate API plan, which is more expensive than a regular subscription.

\paragraph{Overall.}
Models and queries were of the highest quality. They use types and named
constants that ease parameter sweeps, sensible (Danish) parameter values,
well-chosen names and documented declarations, good layout in \uppaal GUI of automata with only few label
overlaps. The increased productivity is substantial: long-standing bugs were fixed in
one session, the end point of earlier work was Claude's first proposal, and
models were built in domains new to the user. Claude also proposed features
and analyses it was not asked for. Trust rested on \uppaal , which refuted
Claude in the Fischer and TSN cases; fidelity to the real system remains for
the domain expert to judge. Each case is a single session, so these are our
judgements rather than measurements.

%% file: conclusions.tex
\uppaai{}\footnote{Available from \url{https://people.cs.aau.dk/~marius/beta/ai}} extends \uppaal{} with a Tools API that lets an LLM construct,
explain, debug and specify models, either from an external agentic client
over MCP or from a chat embedded in \uppaal{}. In the case studies, the
assistant fixed long-standing bugs in a single session, made as its first
proposal the model that had been the end point of our earlier work, built
models in domains new to the user, and proposed features and analyses it had
not been asked for. 

\vspace{\medskipamount}
The sessions also showed where the interface holds the assistant back, and this shapes our future work: verification calls that can be canceled and do not block the editor, so that minute-long checks and learning runs no longer stall the session; access to engine options, the simulator and traces by default; verification of model variants other than the one open in the editor; whole-model replacement that keeps the undo history; and learned strategies
stored with the model. These come in addition to the tools already planned for
exporting templates, describing queries, simulation, and version control
integrated with the chat history.

\vspace{\medskipamount}
Finally, since each case is a single session
with one LLM, we plan to evaluate the embedded chat with the other supported
providers, repeat sessions to assess their variability, and study the use of
\uppaai{} with a wider group of users, building on the session files we have
made publicly available.

%% file: demo.tex
\section{Demo using Anthropic Claude Code with Sonnet 5.5}
\begin{figure}[!tbh]
  \centering
  \includegraphics[width=0.96\textwidth]{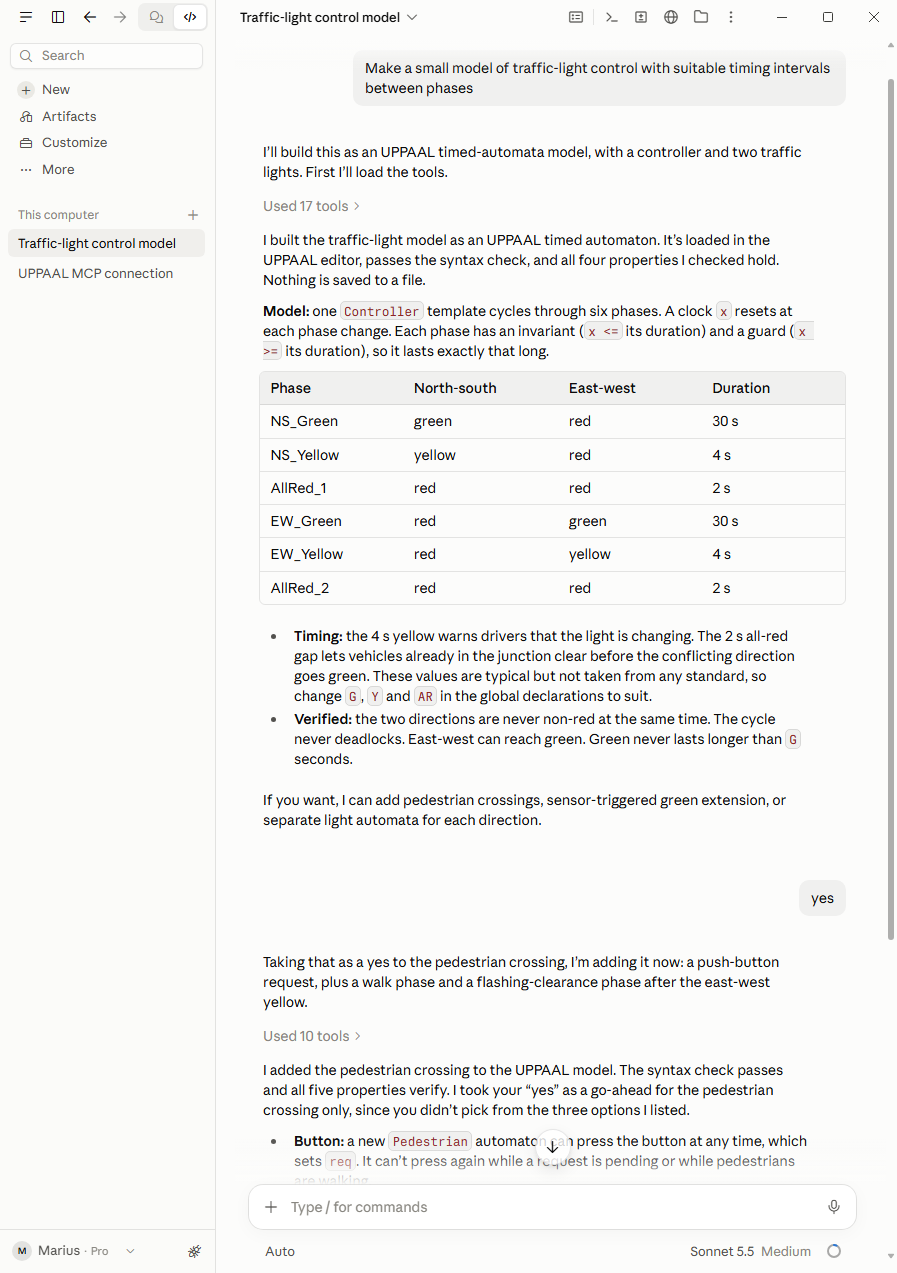}
  \vspace*{-3mm}
  \caption{Claude prompt, result and explanations}\label{fig:ClaudeTraffic}
\end{figure}

\begin{figure}[!h]
  \centering
  \includegraphics[width=0.98\textwidth]{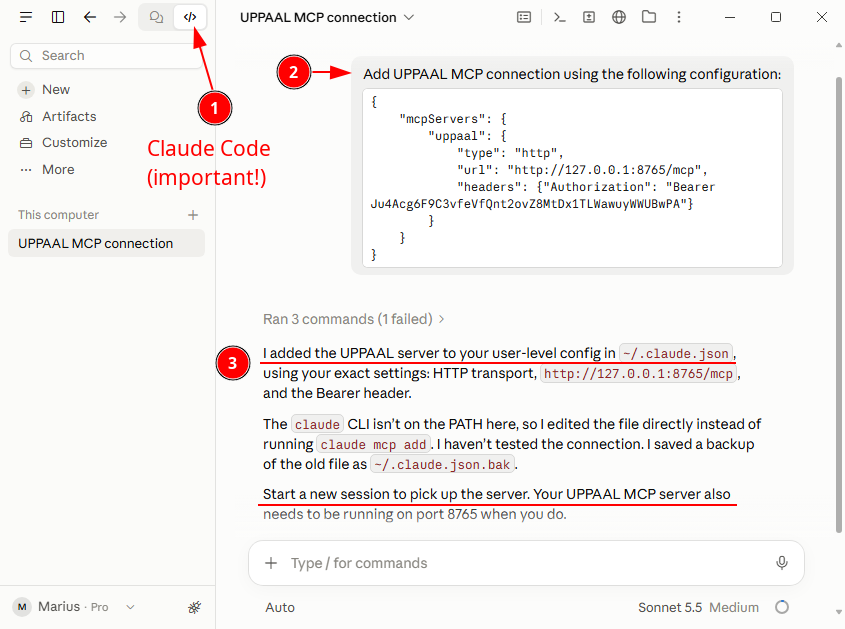}
  \vspace*{-3mm}
  \caption{Configuring \uppaal MCP connection (Fig.\ref{fig:MCPGuide}) within Claude Desktop}\label{fig:ClaudeMCP}
\end{figure}

\begin{figure}[!tbh]
  \centering
  \begin{subfigure}{\textwidth}
    \centering
    \includegraphics[width=\textwidth]{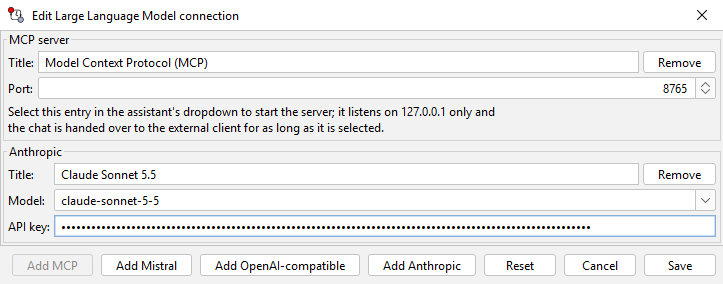}
    \caption{\uppaal dialog for editing LLM connections: \texttt{Edit} $\triangleright$ \texttt{Large Language Models}}\label{fig:EditLLMs}
  \end{subfigure}\bigskip\\
  \begin{subfigure}{\textwidth}
    \centering
    \includegraphics[width=\textwidth]{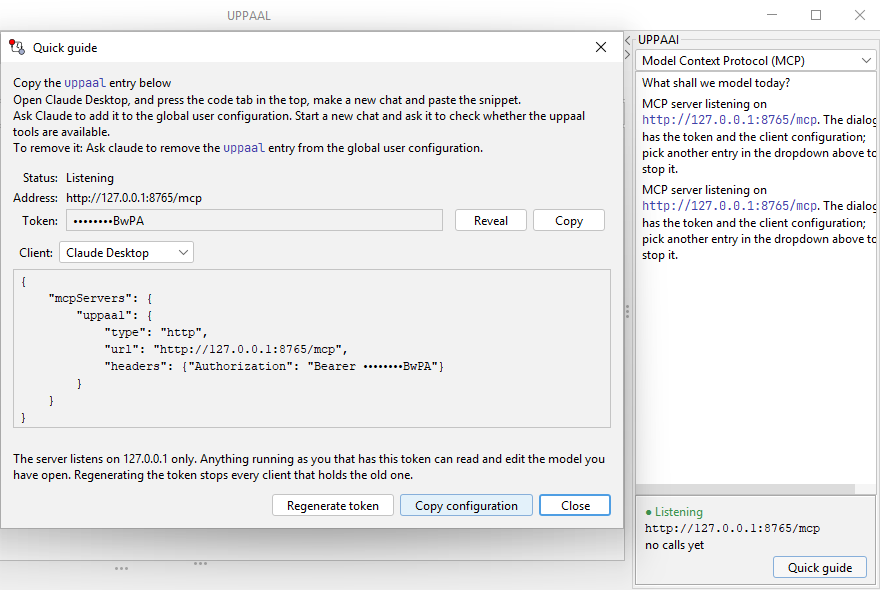}
    \caption{\uppaal MCP configuration: \uppaai $\triangleright$ \texttt{Quick guide} $\triangleright$ \texttt{Copy configuration}}\label{fig:MCPGuide}
  \end{subfigure}
  \caption{LLM configuration dialogs in \uppaal}\label{fig:LLMConfig}
\end{figure}

In Claude Desktop select Claude Code, configure \uppaal MCP (Fig.~\ref{fig:ClaudeMCP},~\ref{fig:LLMConfig}), start a new session and prompt: ``make a small model of traffic control with suitable timing intervals between phases'' (Fig.~\ref{fig:ClaudeTraffic}).
\vspace*{-8mm}

\begin{figure}[!tbh]
  \begin{subfigure}{\textwidth}
    \includegraphics[width=\linewidth]{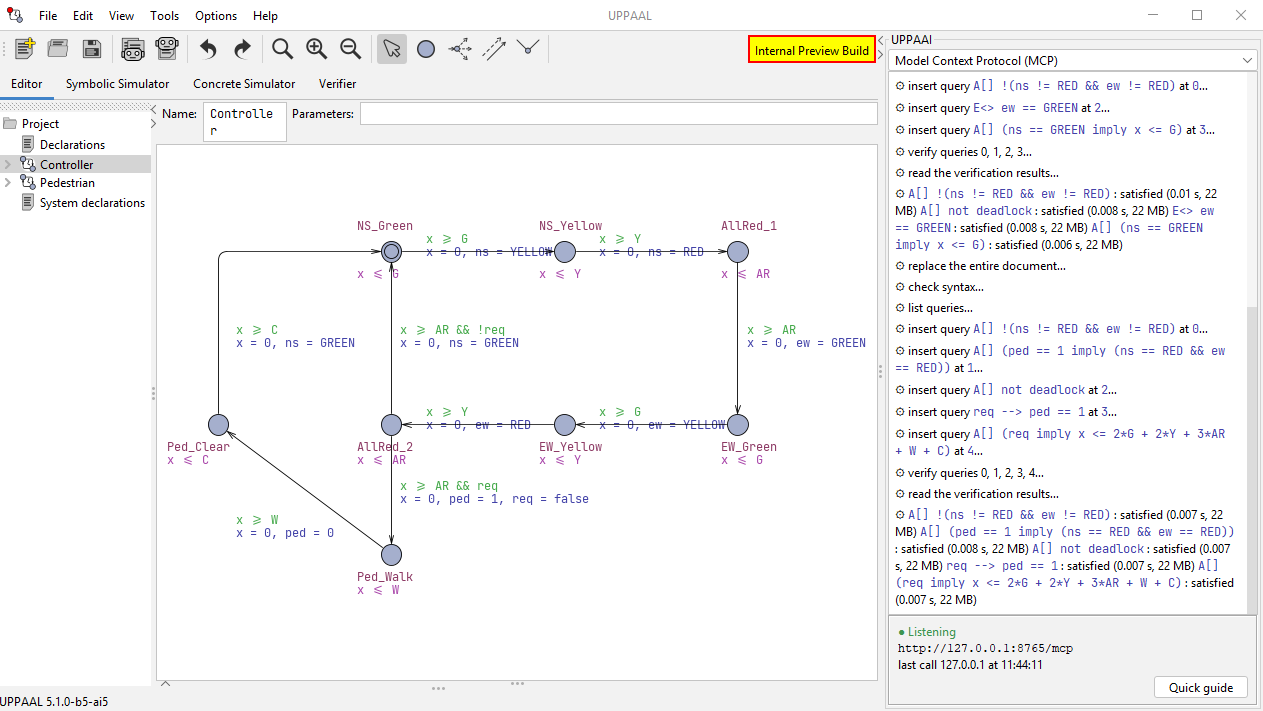}
    \caption{Traffic lights model in \uppaal editor tab}\label{fig:UppaalEditor}
  \end{subfigure}\medskip\\
  \begin{subfigure}{\textwidth}
    \includegraphics[width=\textwidth]{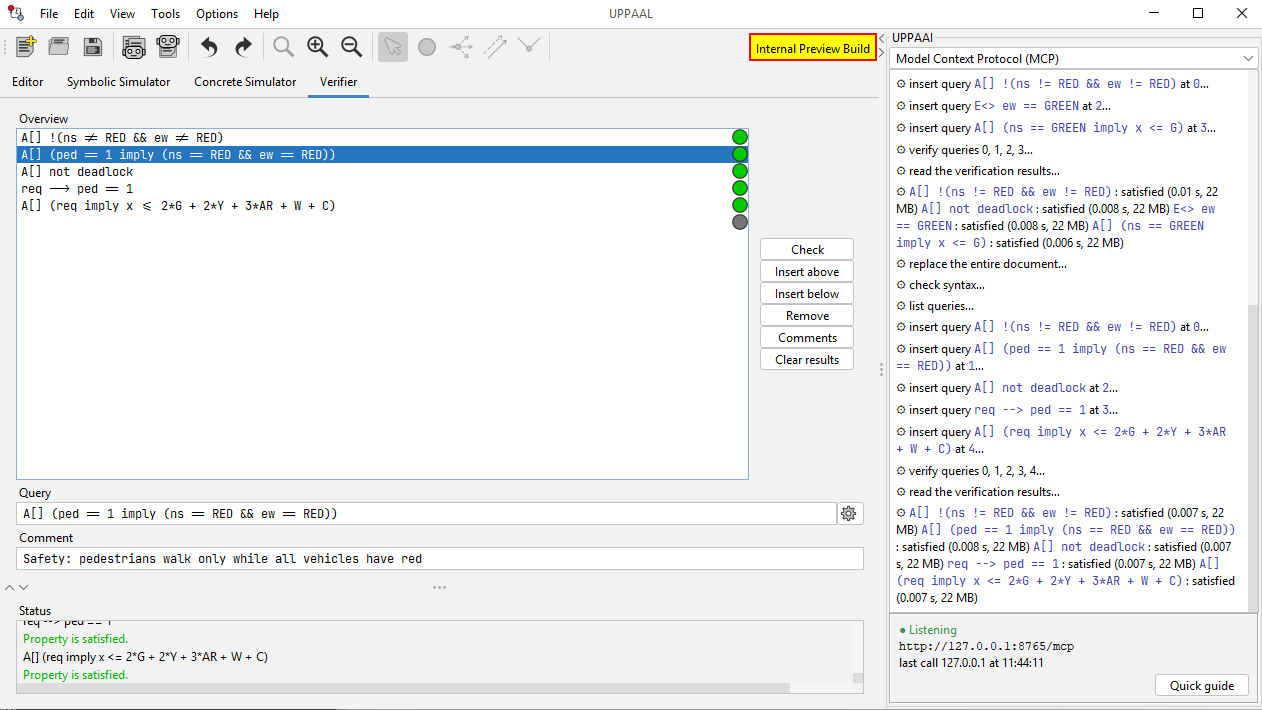}
    \caption{Traffic lights queries in \uppaal verifier tab}\label{fig:UppaalVerifier}
  \end{subfigure}
  \caption{Claude created traffic lights model with queries through \uppaal MCP}
\end{figure}

Observe the model being edited in \uppaal editor tab as Claude works its way through \uppaal Tools API. The history is shown in \uppaai panel on the right (Fig.~\ref{fig:UppaalEditor}). Select \texttt{Declarations} and admire the clean and well structured code with comments.
Once Claude is done, switch to \uppaal verifier tab and inspect the queries Claude wrote and \uppaal checked (Fig~\ref{fig:UppaalVerifier}). Observe that Claude left comments explaining each query.%

\begin{figure}[!h]
  \includegraphics[width=\textwidth]{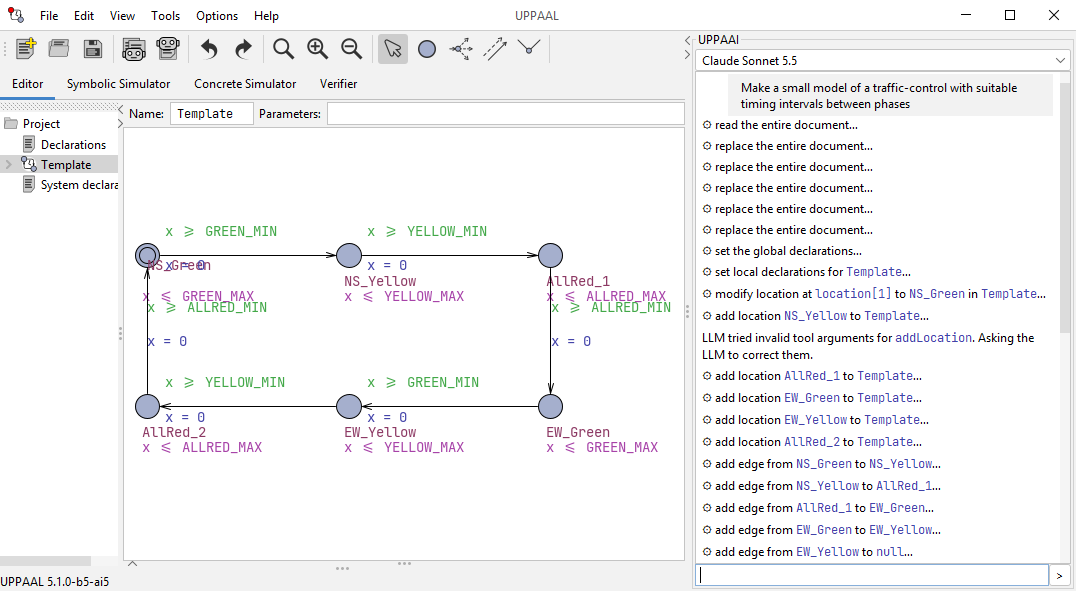}
  \vspace*{-6mm}
  \caption{\uppaal with direct connection to Anthropic API with Claude Sonnet 5.5}\label{fig:UppaalClaude}
\end{figure}

Alternatively, configure \uppaal to use Anthropic API keys with the latest \texttt{claude-sonnet-5-5} model (Fig.~\ref{fig:EditLLMs}), switch to ``Claude Sonnet 5.5'' on \uppaai panel and enter the same prompt into the chat box (Fig.~\ref{fig:UppaalClaude}). The results may vary due to non-deterministic nature of LLMs.%
\vspace*{-4mm}
\vspace*{-12mm}%

%% file: ingredients.tex
\begin{table}[!bth]
\centering\footnotesize
\caption{Ingredients  of a \uppaal{} modeling language.}
\label{tab:ingredients}
\begin{tabularx}{\linewidth}{@{}l>{\ttfamily}lX@{}}
\toprule
\textbf{Ingredient} & \textrm{\textbf{Example syntax}} & \textbf{Role} \\
\midrule
Locations      & \uppLoc{Idle}, \uppLoc{Busy} & Control states. Initial, \emph{urgent} (no delay) and \emph{committed} (atomic, no interleaving) variants \\
Clocks         & \uppStat{clock x, y;} & Real-valued, advancing at rate \uppConst{1}. \emph{Stopwatches} via rate constraints \uppInv{x'==0} \\
Hybrid / ODE dynamics & \uppInv{T' == -k*(T - Tenv)} & Continuous variables with arbitrary rates or ODEs written as derivative constraints in invariants, using \uppType{double} variables and maths functions. These give stochastic \emph{hybrid} automata, analysed by simulation and \smc; the symbolic engines support only stopwatches \\
Invariants     & \uppInv{x <= 5} & Location constraint bounding how long time may pass there \\
Edges          & \uppGuard{g} / \uppSync{sync} / \uppUpdate{upd} & Guard, optional synchronisation and update (clock resets, assignments, function calls), with \uppSelect{select} for families of edges \\
Clock constraints & \uppStat{x - y < 3} & Bounds on clocks and clock differences, in guards and invariants \\
Synchronisation & \uppSync{go!} / \uppSync{go?} & Binary handshake channels, \uppType{broadcast} (one-to-many, non-blocking) and \emph{urgent} channels, and channel arrays \\
Discrete data  & \uppStat{int[0,N] len;} & Bounded integers, booleans, constants, arrays, structs and \uppType{typedef}s, with strong type and range checking \\
User functions & \uppStat{void enqueue(id_t e)} & C-like functions with local scope, loops and ranged \uppStat{forall}/\uppStat{exists}/\uppStat{sum} \\
Templates      & \uppStat{Train(const id_t id)} & Parameterised automata. Partial and automatic instantiation over bounded ranges in the \texttt{system} declaration \\
Stochastic annotations & \uppRate{rate}, \uppWeight{weights} & Exponential rates for unbounded delays (uniform delays otherwise) and probabilistic weights on branches \\
Game annotations & \textrm{solid / dashed} & Controllable vs.\ uncontrollable edges \\
\bottomrule
\end{tabularx}
\end{table}

%% file: queries.tex
\begin{table}[!bth]
\centering\footnotesize
\caption{Formalisms, queries and results supported by \uppaal{} 5.0.}
\label{tab:formalisms}
\begin{tabularx}{\linewidth}{@{}>{\raggedright}p{2.1cm}>{\raggedright}p{2.6cm}>{\ttfamily\raggedright}p{4.1cm}X@{}}
\toprule
\textbf{Formalism} & \textbf{Non-determinism} & \textrm{\textbf{Typical queries}} & \textbf{Results produced} \\
\midrule
Timed automata (symbolic MC) & Non-deterministic: all behaviours are explored exhaustively &
\uppProp{E<> p}, \uppProp{A[] p}, \uppProp{E[] p}, \uppProp{A<> p}, \uppProp{p --> q}, \uppProp{A[] not deadlock}, \uppProp{sup}/\uppProp{inf}/\uppProp{bounds\{p\}: e} &
Boolean verdict (``maybe'' for stopwatch over-approximation). Symbolic \emph{diagnostic trace} (witness or counterexample). Upper and lower bounds \\
\addlinespace
Stochastic (hybrid) TA (SMC) & Probabilistic: uniform or exponential delays, weighted branches, race between components &
\uppProp{Pr[<=T](<> p)}, \uppProp{Pr[<=T](<> p) >= 0.9}, \uppProp{Pr[..](..) >= Pr[..](..)}, \uppProp{E[<=T;n](max: e)}, \uppProp{simulate[<=T;n]\{e1,e2\}} &
Probability estimates with confidence intervals. Hypothesis-test and comparison verdicts. Expected values with confidence intervals. Sampled trajectories, histograms and concrete witness traces \\
\addlinespace
Timed games (TIGA) & Antagonistic environment controls the uncontrollable edges &
\uppProp{control: A[] p}, \uppProp{control: A<> p} &
Verdict, plus a synthesised \emph{most permissive} winning strategy that is correct by construction \\
\addlinespace
Timed MDPs (Stratego) & Stochastic environment, controller chooses &
\uppProp{strategy S = minE(e)[<=T]: <> p under Safe}, \uppProp{Pr[..](<> p) under S}, \uppProp{E[..](max: e) under S}, \uppProp{simulate[..]\{..\} under S} &
Near-optimal strategies \emph{learned} by reinforcement learning, optionally shielded by a synthesised safe strategy. Strategies can be evaluated, compared and model checked \\
\bottomrule
\end{tabularx}
\end{table}